\documentclass[prb,reprint]{revtex4-1}

\usepackage{graphicx}
\usepackage{multirow}

\begin{document}

\title{Ab-Initio calculations of binding energy of In and Ga adatoms on three GaAs(111)A surface reconstructions}
\author{J. N. Shapiro}
\author{D. L. Huffaker}
\affiliation{California Nanoscience Institute and Electrical Engineering Dept., University of California at Los Angeles, Los Angeles Ca. 90095}
\author{C. Ratsch}
\affiliation{Institute for Pure and Applied Mathematics and Dept. of Mathematics, University of California at Los Angeles, Los Angeles, CA 90095}

\date{\today}

\begin{abstract}
Calculations of the potential energy surface for tracer Ga and In adatoms above three GaAs (111)A surface reconstructions are presented in order to understand the growth conditions required to form axial heterostructures in GaAs/InGaAs nano-pillars.  In all calculations the Ga adatom has a stronger bond energy to the surface than the In adatom.  The diffusion barriers for Ga adatoms are 140meV larger than for In adatoms on the Ga vacancy surface, but they are comparable on the As trimer surface. Also the binding energy for an In adatom is closer to that of a Ga adatom on the As trimer surface.  We conclude that the As trimer surface is preferable for adsorption of In and thus for selective formation of hetero-interfaces on (111) facets.  This work helps explain the recent successful formation of axial GaAs/InGaAs hetero-interfaces in catalyst free nano-pillars.
\end{abstract}

\maketitle

\section*{Introduction}

Semiconductor nanowires(NWs) and nanopillars(NPs) are exciting  materials for
probing mesoscopic physics and as building  blocks for future high performance opto-electronic devices on Si\cite{Fuhrer2007Few,Xiang2006,Lu:2005lr}. NP synthesis by catalyst-free selective area metal-organic chemical vapor deposition (SA-MOCVD) is a growth technique for forming large arrays of uniform NPs in lithographically defined locations with the inclusion of optical alignment marks for device integration \cite{Akabori:2003}.  

The absence of a metal particle to catalyze growth means that atoms adsorb directly onto the crystal surfaces from the vapor, and the resulting crystal shape is controlled in part by minimization of the total surface energy\cite{ikejiri:2008}.  GaAs nanopillars grow in the [111] direction, and have hexagonal symmetry with side facets composed of the $\{01\bar{1}\}$ family of planes. Atoms from the vapor adsorb on all facets of the NP and then diffuse to the (111) surface at the tip where they incorporate.  The polar (111) surface has a higher surface energy than the stoichiometric $\{01\bar{1}\}$ planes, making the observed crystal shape energetically favorable. 
	
\begin{figure}
	\includegraphics[scale=0.4]{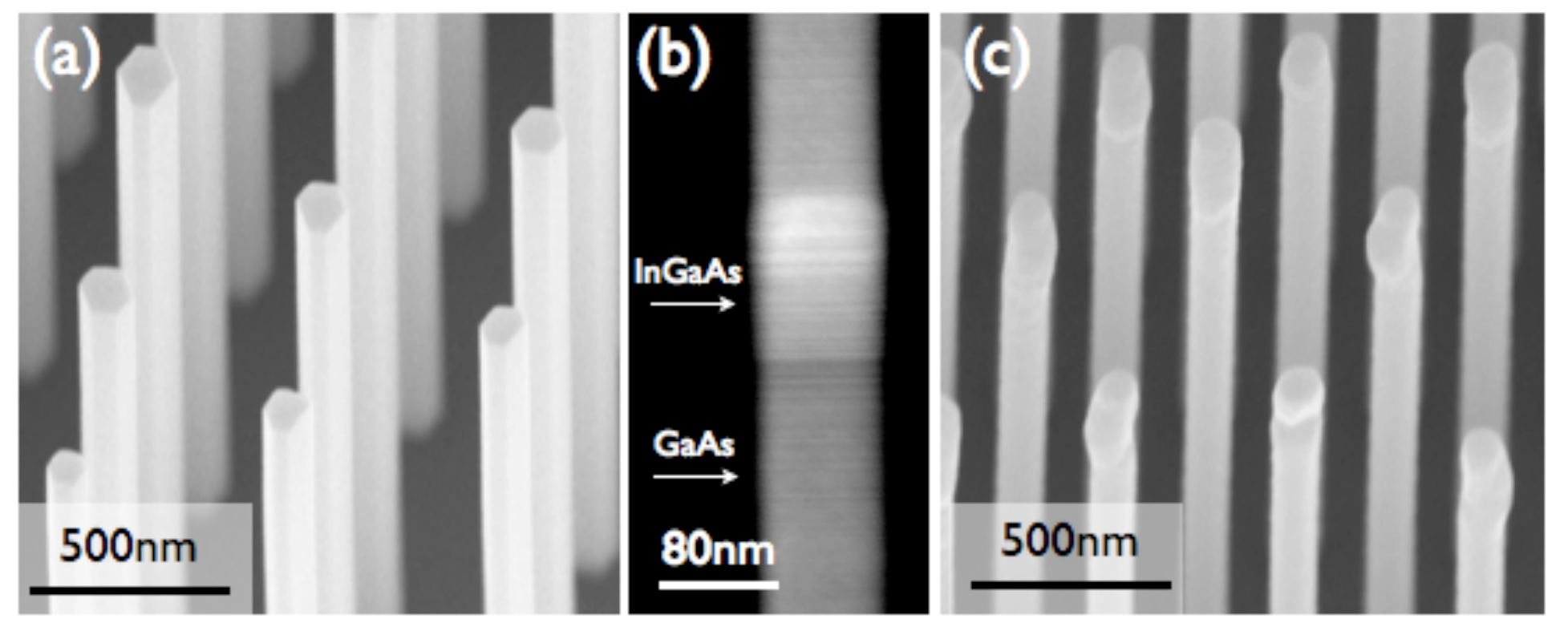}
	\caption{\label{fig:pillar_image}(a) SEM of GaAs nanopillars containing axial InGaAs inserts grown at high V/III ratio. (b) Dark field STEM of single InGaAs insert. (c) SEM of GaAs nanopillars terminated with InGaAs at low V/III ratio.}
\end{figure}

Heterostructure formation is a necessary capability to master in catalyst-free NP synthesis in order to create efficient optical devices\cite{Agarwal:2008lr}.  Core-shell hetero-structures have been studied in a variety of material systems, but axial hetero-structure formation has been elusive in this growth mode. When a new atomic species is introduced, the surface energetics must promote incorporation of the new species on the top (111) surface while simultaneously suppressing incorporation on the side walls.  Despite this challenge, axial InGaAs segments of varying composition and thickness were recently demonstrated in GaAs catalyst free NPs grown by SA-MOCVD\cite{shapiro:2010a}.  High V/III ratios ($V/III\sim50$) were required to promote incorporation of In in the axial direction with negligible shell growth.  At the lower V/III ratios ($V/III\sim10$) typically used for GaAs NP homoepitaxy, indium is not selective to the (111) surface, and instead nucleates on the side-walls, deforming the crystal facets.  Fig.~\ref{fig:pillar_image}(a) shows scanning electron micrograph (SEM) of NPs formed by SA-MOCVD with axial InGaAs inserts at high V/III ratio, and the vertical side-walls and hexagonal symmetry are evident.  Fig.~\ref{fig:pillar_image}(b) shows a dark field scanning transmission electron micrograph (STEM) of the same pillars revealing the axial InGaAs segment.  In contrast,  Fig.~\ref{fig:pillar_image}(c) shows pillars terminated with InGaAs sections at low V/III ratios, and having deformed crystal facets due to indium nucleation on the side-walls.  This tendency for indium to bond to all available crystal surfaces has also been reported in Ref~[\onlinecite{Paetzelt2008Selectivearea}]. 

To investigate possible reasons for the observed differences in behavior between In and Ga during nanopillar epitaxy,
we present a theoretical investigation of the potential energy surface (PES) for Ga and In tracer adatoms situated above three common surface
reconstructions of GaAs(111)A.   The technique of calculating a PES has been applied by numerous researchers as a tool for studying diffusion, adsorption and desorption and for understanding epitaxy on crystal surfaces\cite{Taguchi2000Firstprinciples,Taguchi1999Stable,Penev2004Anisotropic}.  A similar study of In and Ga tracer diffusion on GaAs $\{01\bar{1}\}$ is necessary for a more complete understanding of NP epitaxy, and will be presented in a future publication.  Computational methods are discussed first, followed by a description of the calculations and their results. We conclude with a discussion and interpretation of the results.

\section*{Computational Methods}

\begin{figure*}	\includegraphics[scale=1.0]{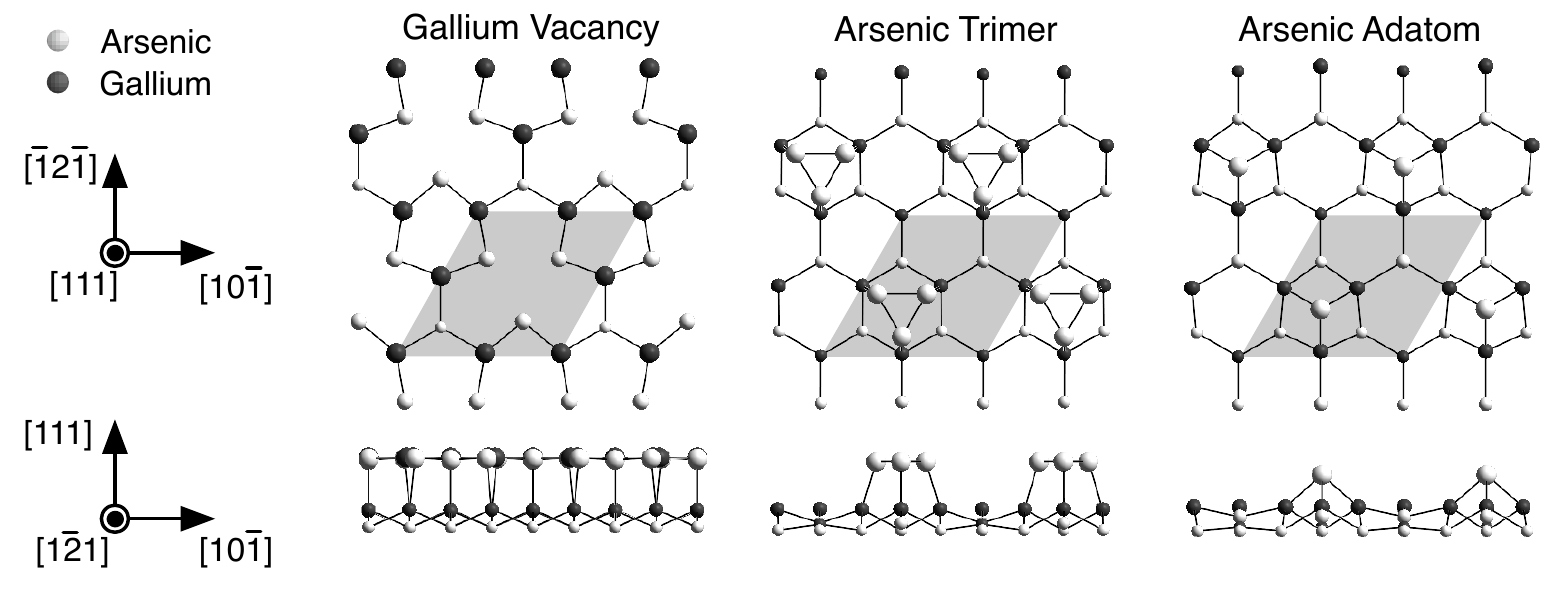}
\caption{\label{fig:surface_reconstructions}Surface reconstructions of the GaAs(111)A surface.  From left to right the Ga vacancy surface, the As Trimer surface and the As adatom surface. Arsenic atoms are depicted as light gray spheres, and gallium atoms are depicted as dark gray spheres. Top and side views are rendered with two or three layers of atoms  The atomic diameters are drawn larger for atoms closer to the surface. The 2x2 unit cell is identified by a shaded parallelogram.  While all surfaces possess a 2x2 unit cell, the PES calculations were performed using a 4x4 super-cell to suppress interaction effects between the adatoms.}		
\end{figure*}

To calculate the potential energy surface (PES) of a Ga or In adatom above a GaAs(111)A surface reconstruction, we begin by computing the equilibrium surface geometry of the three reconstructions depicted in Fig.~\ref{fig:surface_reconstructions}.  From left to right, the surfaces are the Ga vacancy surface, the As trimer surface and the As adatom surface.  All three surfaces possess a 2x2 unit cell indicated by a shaded parallelogram. Slabs 9 mono-layers thick are iteratively relaxed, keeping the bottom three mono-layers fixed, until residual atomic forces are $<0.02$eV/\AA.  

The total energy of the surface with an additional Ga or In adatom is then computed using a 4x4 super cell.   The entire system, slab and adatom, is allowed to relax, but  the adatom coordinates are fixed perpendicular to the [111] direction (the adatom is fixed in the x-y plane and allowed to relax in z).  All three surfaces possess 3-fold rotational symmetry, and each rotationally symmetric slice posses a mirror symmetry such that only 6-8 points are sampled in a triangle above the 2x2 unit cell.  The calculated energies are then reflected, rotated twice through $120^\circ$ and mapped to a rectilinear grid using a cubic interpolation to generate a PES for the adatom of interest.  The energy zero-point is chosen to be the total energy of the relaxed reconstructed surface plus the total energy of an isolated atom of In or Ga. 

Calculations were performed within the framework of density-functional theory (DFT) as implemented in the software package FHI-AIMS\cite{Blum20092175}, which uses numeric atom centered orbitals for its basis set. The Perdew-Burke-Ernzerhof (PBE) parameterization of the generalized gradient approximation is used for the exchange correlation functional\cite{PBEGGA}.  Approximately 16 layers of vacuum and 64 equivalent k-points in the 1x1 unit cell are specified.  Convergence of the energy difference between the maximum and minimum on the PES is confirmed for the k-points, slab thickness, vacuum layers and super-cell size for the Ga vacancy reconstruction.  In addition, total energy differences were tested for the FHI-AIMS built in settings ``light'' and ``tight''~\cite{Blum20092175}.  The ``light'' setting, having fewer basis functions and a smaller numerical integration mesh, is found to increase the speed of the calculation while generating results that differ from ``tight'' by only a few meV.  Calculations are therefore carried out using the ``light'' setting. 

\section*{Results}

The potential energy surfaces for indium and gallium adatoms above each surface reconstruction are presented in this section.  The binding energies at adsorption sites, $A_i$, and transition points, $T$ and $T^\prime$, for In and Ga above each surface are collected in Table~\ref{tab:summary}. The dominant diffusion energy barriers, calculated as the difference $E_D = T - A_1$, are also tabulated.  The Ga vacancy and the As trimer surfaces are the primary surfaces of interest because they are energetically favorable in a vapor consisting of mixed As and Ga atoms\cite{PhysRevB.54.8844}.  The Ga vacancy surface has the lower surface energy at low As chemical potentials and the As trimer surface as the lower surface energy at high As chemical potentials. The As adatom reconstruction always has the highest relative surface energy, and is presented here for completeness, even though this  surface does not exist with high probability. 

\begin{table}
	\caption{\label{tab:summary}Diffusion barrier $E_D$, minimum potential energy $A_1$, secondary minimum potential energy $A_2$ and transition point $T$ of In and Ga adatoms above three common surface reconstructions of GaAs (111)A.  All values are in electron volts(eV).}
	\begin{ruledtabular}
	\begin{tabular}{ccccccc}
	Surface & Adatom &  $E_D$ & \textbf{$A_1$} & \textbf{$A_2$} & \textbf{$T$}  & \textbf{$T^\prime$} \\
\hline
	\multirow{2}{*}{Ga Vacancy} & Ga & \textbf{1.06} & -2.87 & -2.21 & -1.81 & -1.73  \\
	 & In & \textbf{0.92} & -2.65 & -2.06 & -1.73 & -1.65  \\
\hline
	\multirow{2}{*}{As Trimer} & Ga & \textbf{0.27} & -7.10 & - & -6.83 & - \\
	 & In & \textbf{0.26} & -6.99 & -6.88 & -6.73 & - \\
\hline
	\multirow{2}{*}{As Adatom} & Ga & \textbf{0.20} & -6.88 & - & -6.68 & - \\
	 & In & \textbf{0.13} & -6.71 & - & -6.58 & -\\
	\end{tabular}
\end{ruledtabular}
\end{table}

\begin{figure}
\includegraphics[scale=0.9]{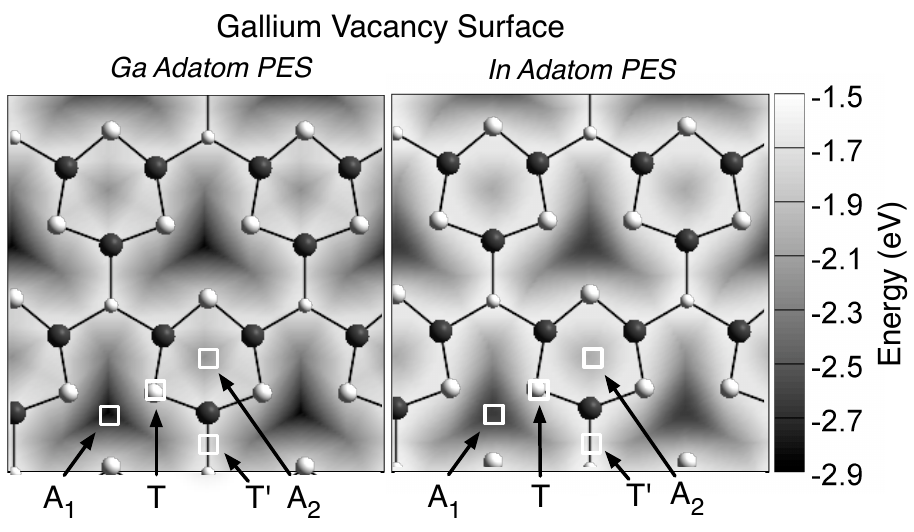}		
\caption{\label{fig:GaVac_PES}Gallium vacancy potential energy (PES) surface for Ga (left) and In(right) adatoms.  The top atomic layer of the reconstruction is drawn as an overlay to assist in visualizing the adsorption sites.  The primary potential minimum occurs at $A_1$ with a secondary minimum at $A_2$.   At typical NP growth temperatures, diffusion is dominated by hops between $A_1$ and $A_2$ over the saddle points labeled $T$.}
\end{figure}

The PES for Ga and In adatoms above the Ga vacancy surface reconstruction are shown in Fig.~\ref{fig:GaVac_PES}.  This surface, characterized by a missing Ga atom, is thought to be the stable reconstruction under most conditions except in extremely arsenic rich environments\cite{PhysRevB.54.8844}.  Comparing In and Ga adatoms above the Ga vacancy surface, the PES are qualitatively similar, with a deep minimum at the vacancy site $A_1$, and a secondary minimum at the site $A_2$, above third layer As atoms.  The transition points, $T$ and $T^\prime$, are saddle points of the PES that are crossed when hopping between adsorption sites. 

Diffusion can occur on the Ga vacancy surface by two possible pathways.  Either the adatom hops directly between $A_1$ sites over the transition point $T^\prime$, or it crosses over the point $T$ into the secondary site $A_2$, and then rapidly hops back into an adjacent $A_1$ site. The energy barrier $T^\prime - A_1$ is 80 meV larger than the transition over $T$.  At typical growth temperatures of $~\sim1000\;K$, diffusion between $A_1$ sites by way of $A_2$ is fast enough to dominate the diffusion path.  The diffusion barrier, $E_D$, reported in table~\ref{tab:summary} is the barrier to hop from $A_1$ to $A_2$.  

Ga atoms are less mobile than In on this surface with a diffusion barrier 140 meV higher than In regardless of the path taken.  The binding energy of a Ga adatom at $A_1$ is 220 meV larger than for In, suggesting that Ga adatoms will be adsorbed preferentially over In adatoms.  This calculation agrees with the observation that In floats to the surface when forming the NP hetero-interface\cite{shapiro:2010a}.

The PES for a Ga adatom above the Ga vacancy reconstruction was previously calculated by Taguchi \textit{et. al.}  in Ref.~[\onlinecite{Taguchi1999Stable}]; however, our results are significantly different.  In that work, contrary to expectations, they found the potential energy minimum was not in the lattice site vacated by the Ga atom, but at adjacent interstitial locations with diffusion energy barriers of $\sim0.4$eV.  Our calculations, in contrast, show a deep potential minimum at the vacant lattice site with diffusion barriers $\sim1.0\text{eV}$.  We are unable to explain the discrepancies between the two calculations, however, the authors of Ref~[\onlinecite{Taguchi1999Stable}] acknowledge that the vacant Ga site should be more stable according to calculations based on the interatomic potential.  In light of the conflicting results, we carefully checked our energy calculations and algorithms for generating the PES, which exploit the surface symmetry, and are unable to find errors in our methods.

\begin{figure}
\includegraphics[scale=0.9]{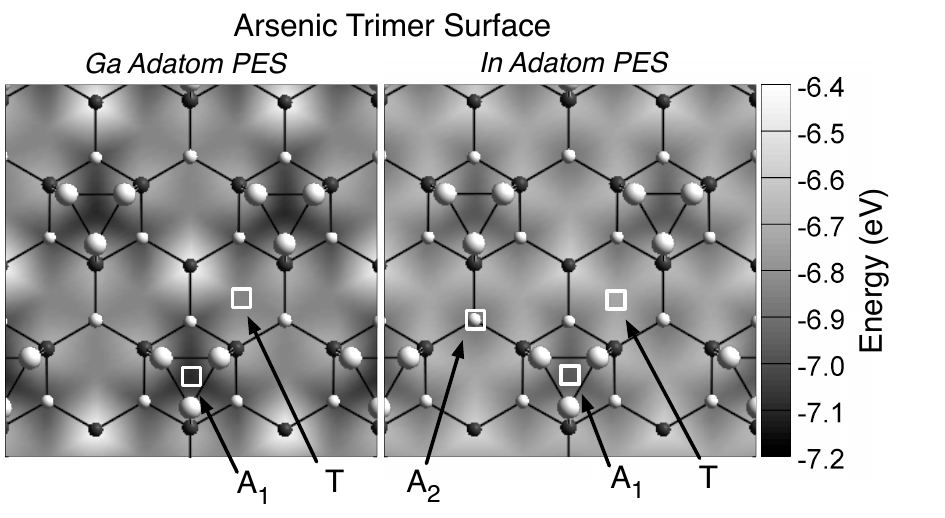}		
\caption{\label{fig:AsTrimer_PES}Potential energy surface for Ga (left) and In (right) adatoms above the As Trimer reconstruction of GaAs(111). The top atomic layers of the reconstruction are drawn as an overlay to assist in visualizing the adsorption sites.}
\end{figure}

The As trimer PES for In and Ga adatoms are presented in Fig.~\ref{fig:AsTrimer_PES}.  The As trimer surface is the stable reconstruction appearing in arsenic rich environments.  The PES for both In and Ga adatoms have potential energy minimum $A_1$ at the center of the As trimer, and a diffusion barrier height of 260-270 meV. The PES for an In adatom also has a secondary minimum, $A_2$, above one of the second layer As atoms that can potentially slow the diffusion for In.  The difference in binding energy between Ga and In is only 110 meV for the As trimer surface, compared to 220 meV for the Ga vacancy surface. Indium adatoms will have a higher probability of adsorption on this surface compared to the Ga vacancy surface because of the equivalent diffusion coefficients and more competitive binding energy.

\begin{figure}
\includegraphics[scale=0.9]{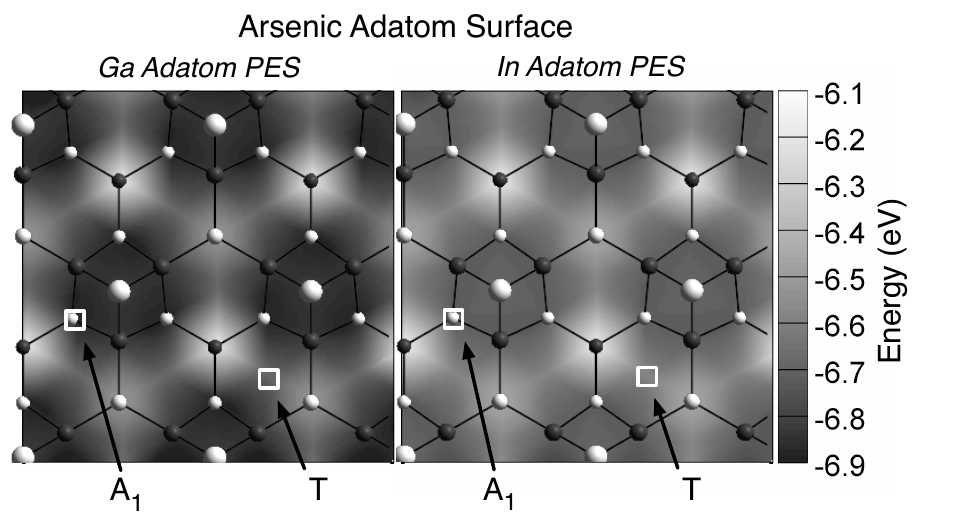}		
\caption{\label{fig:AsAdatom_PES}Potential energy surface for Ga (left) and In(right) adatoms above the As Adatom reconstruction of GaAs(111).}
\end{figure}

The PES for the In and Ga adatoms above the As adatom surface are shown in Fig.~\ref{fig:AsAdatom_PES}.  The stable adsorption sites, $A_1$, are above the top layer As atoms, but the entire region surrounding the As adatom is energetically very flat with minor variations of tens of meV.  The region surrounding the As adatom is therefore an adsorption site.   Diffusion barriers are lowest for this surface reconstruction compared to the other surface reconstructions with a barrier of 130 meV for In and 200 meV for Ga.  The absolute binding energy is  170meV larger for Ga than for In on the As adatom surface reconstruction.

\section*{Discussion}

Calculations were performed to provide a physical explanation for why high V/III ratios are required for formation of GaAs/InGaAs axial heterostructures in (111) oriented catalyst free NPs.  We believe that the high As chemical potential results in surface reconstructions on the NP that suppress In incorporation on the $\{01\bar{1}\}$ NP side-walls while simultaneously promoting In incorporation on the (111) NP tip. 

The  calculations reported in this work support the hypothesis that the As trimer surface, stable at high As chemical potential, is desirable for higher rates of indium incorporation on the (111) surface for two reasons. First, the difference in binding energy between Ga and In adatoms in the $A_1$ adsorption site is reduced from 220meV on the Ga vacancy surface to 110meV on the As trimer surface.  This reduction means that In adatoms compete more effectively with Ga and have a higher probability incorporating into the crystal in the presence of the As trimer surface.  Second the diffusion barriers, $E_D$ are comparable for both Ga and In adatoms on the As trimer surface, yet the diffusion coefficient of In is roughly two orders of magnitude larger on the Ga vacancy surface at typical growth temperatures of $\sim1000 K$.  In the presence of a Ga vacancy surface, In adatoms will diffuse more quickly than Ga and desorb from the small (111) surface at the tip of the pillar.  The resulting chemical environment of adsorbates at the pillar tip will be richer in Ga than in the surrounding vapor.  In contrast, the comparable diffusion barriers of both Ga and In on the As trimer surface  will result in a concentration of adsorbates representative of the concentration in the surrounding vapor.  The two reasons cited explain why In adatoms incorporate more efficiently on the (111) surface at high As chemical potential. 

In summary, calculations of the PES for tracer In and Ga adatoms above three stable surface reconstructions of GaAs(111)A were performed.  Gallium has a larger binding energy to all surfaces, yet the binding energy of In is  competitive with Ga on the As trimer surface.  In addition, the diffusion barriers for In and Ga are identical on the As trimer surface, which forms at high As chemical potentials.  On the Ga vacancy surface, stable at low As chemical potentials, indium is calculated to rapidly diffuse, lowering the residence times at adsorption sites, reducing the opportunity for incorporation into the crystal.  These results suggest that formation of an As trimer surface at high V/III ratios promotes formation of axial GaAs/InGaAs hetero-interfaces during nanopillar growth.

The authors gratefully acknowledge the financial support of NSF (Grant no. DMR-1007051 and  DMS-0439872), AFOSR (Grant no. FA9550-08-1-0198), DOD (Grant No. NSSEFF N00244-09-1-0091), and the NSF Clean-Green IGERT Fellowship funding for Mr. Shapiro (Grant no. DGE-0903720). 

\bibliography{GaAs111}
\end{document}